\documentclass[aps,pre,twocolumn,float]{revtex4}
\usepackage{amsmath}
\usepackage{amsfonts}
\usepackage{amssymb}
\usepackage{upgreek}
\usepackage{amsmath,bm,epsfig}

\usepackage{subfigure}

\def\a{\upalpha}
\def\O{\mathcal{O}}
\def\e{\varepsilon}
\def\Integer{\mathbb{Z}}
\newcommand{\ue}{\mathrm{e}}
\begin{document}

\title{Observation of the Inverse Energy Cascade\\ in the modified Korteweg--de Vries Equation}
\author{D. Dutykh$^{\dag}$ and E. Tobisch}
 \email{Elena.Tobisch@jku.at}
  \affiliation{$^\dag$ LAMA, UMR 5127 CNRS, Universit\'e de Savoie, Campus Scientifique, 73376 Le Bourget-du-Lac Cedex, France\\
$^*$Institute for Analysis, J. Kepler UNiversity, Linz; Austria}

\begin{abstract}
In this Letter we demonstrate for the first time the formation of the inverse energy cascade in the focusing modified  Korteweg--de Vries (mKdV) equation. We study numerically the properties of this cascade such as the dependence of the spectrum shape on the initial excitation parameter (amplitude), perturbation magnitude and the size of the spectral domain. Most importantly we found that the inverse cascade is always accompanied by the direct one and they both form a very stable quasi-stationary structure in the Fourier space in the spirit of the FPU-like reoccurrence phenomenon. The formation of this structure is intrinsically related to the development of the nonlinear stage of the Modulational Instability (MI). These results can be used in several fields such as the internal gravity water waves, ion-acoustic waves in plasmas and others.
\end{abstract}


\maketitle

\section{Introduction}

Currently there exist three main models which describe the energy cascade in nonlinear wave systems. Each model covers its own range of the nonlinearity parameter. Namely, for strong nonlinearities the Kolmogorov cascade with the widely known exponent $k^{-5/3}$ ($k$ being the vortex scale) introduced for the hydrodynamic turbulence theory \cite{Kolm1}. In the opposite limit of weak nonlinearities, the kinetic weak turbulence spectra $k^{-\alpha}$ (now $k$ is the wavenumber) were introduced, \cite{ZLV}; in this case the spectrum exponent $\alpha$ is not universal anymore and the nonlinearity parameter $ak \lesssim 10^{-2}$, where $a$ is the typical wave amplitude. For the intermediate values of the nonlinearity $ak\ =\ \O\bigl(10^{-1}\bigr)$ the Dynamic energy cascade ($D$-cascade) model was proposed for the equations which feature the Modulational Instability (MI) property \cite{Kart1}. This construction was originally described and studied in details for the NLS-type equations \cite{Kart2, Kart3}.

\begin{figure}
  \includegraphics[width=7cm,height=6cm]{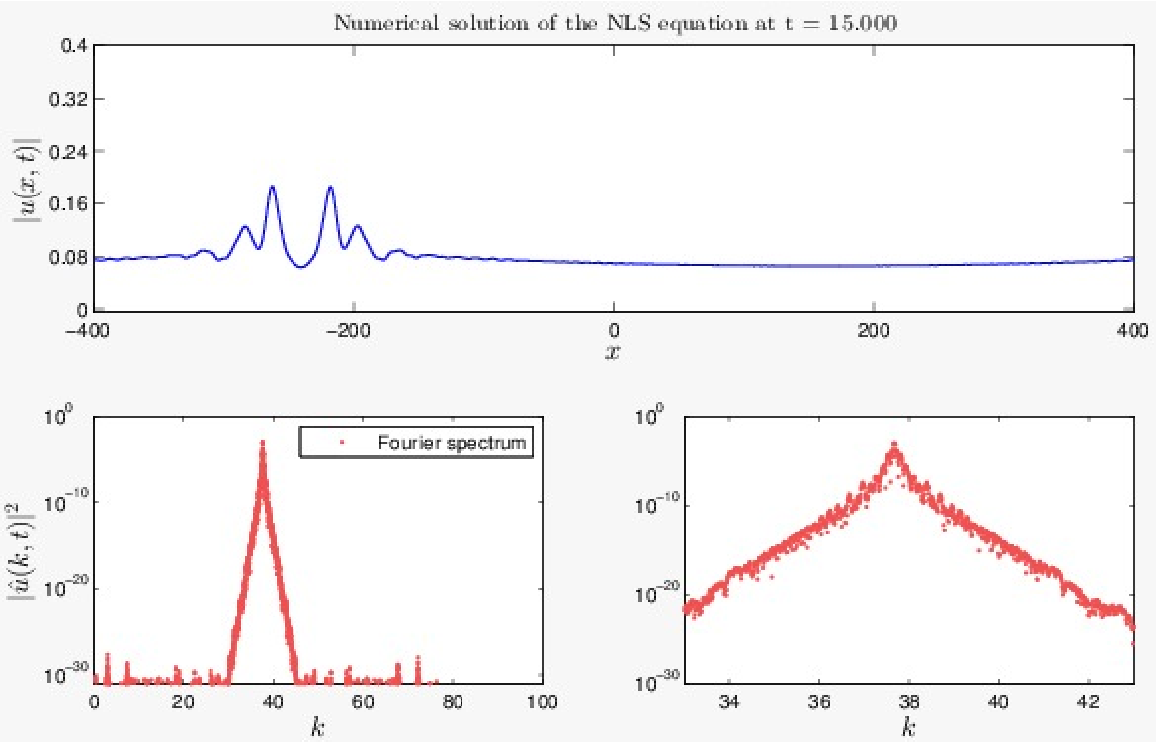}
  \caption{\small\em Modulational instability in the NLS equation.}
  \label{fig:nls}
\end{figure}

In particular it was shown \cite{Kart1} that in the focusing (\textit{i.e.} possessing the MI) NLS equation only one cascading mode is forming due to the intrinsic narrow band spectrum approximation. This prediction was checked numerically and the development of the MI in the NLS is shown on Figure~\ref{fig:nls}. On this bottom panel one can see only one spike in the Fourier spectrum along with the broadening wings.

On the other hand, the modified Korteweg--de Vries (mKdV) equation does not have any restrictions on the wave spectrum. In this study we consider the following focusing mKdV equation:
\begin{equation}\label{mKdV}
  \mathrm{mKdV}(u) \doteq u_t + u_{xxx} + 6u^2u_x = 0\ .
\end{equation}
The formation of the direct $D$-cascade in this equation has been recently observed and investigated numerically in \cite{DT14a}. The direct cascade forms starting from any base wavenumber $k_0 > 0$. The main properties of the direct cascade will be briefly described in the next Section. Concerning the inverse cascade, it was not observed yet to our knowledge. In this Letter we aim at presenting solid numerical evidences towards the existence of the inverse $D$-cascade and studying some of its properties comparing to the direct one.

\section{Direct cascade in the mKdV equation}

For our numerical simulations we adopt the set-up used also earlier in \cite{Grimshaw2001, Grimshaw2005} for the study of the modulation instability in the physical space, with the following initial condition posed on a periodic domain $[-\ell, \ell]\ \doteq\ \bigl[-{\pi}/{k_0}, {\pi}/{k_0}\bigr]$:
\begin{equation}\label{eq:pert}
  u(x,0) \equiv u_0(x) = a\bigl(1 + \delta\sin(K_0 x)\bigr)\sin(k_0 x),
\end{equation}
where $a$ is the base wave amplitude, $\delta$ is the perturbation magnitude and the wavenumbers $k_0$, $K_0$ are chosen such that their ratio $k_0/K_0\in\Integer$. It is important to realize that the amplitude $a$ is the measure of the nonlinearity in the mKdV equation, in the contrast to the NLS equation where the nonlinearity is related to the wave steepness $a K_0$ defined in terms of the modulational wavenumber $K_0$.

\begin{table}
  \centering
  \begin{tabular}{c|c}
  \hline\hline
  \textit{Parameter} & \textit{Range} \\
  \hline\hline
  Amplitude & $a\ =\ 10^{-2}\ \div\ 2.4\times 10^{-1}$ \\
  Perturbation magnitude & $\delta\ =\ 5\times 10^{-2}\ \div\ 5\times 10^{-1}$ \\
  Base wavenumber & $k_0\ =\ 1.8\ \div\ 60.0$ \\
  Simulation time horizon & $T\ =\ 40\ \div\ 2000$ \\
  Fourier harmonics & $N\ =\ 1024\ \div\ 131\, 072$ \\
  \hline\hline
  \end{tabular}
  \bigskip
  \caption{\small\em Parameters range used in numerical simulations.}
  \label{tab:params}
\end{table}

The main characteristics of the direct $D$-cascade observed in \cite{DT14a} are as follows:
\begin{itemize}
  \item (\textit{a}) the main structure of the cascade is already observable in Fourier spectra from the first instances of the dynamical evolution;
  \item (\textit{b}) the development of the Modulation Instability (MI) in the physical space occurs at the time scale $t\ \sim\ \e^{-2}$ and corresponds to the spectral broadening of the cascading modes; the frequencies and energies of the cascading modes are quasi-stationary;
  \item (\textit{c}) the energy spectrum of the cascade has exponential form, $E_k\ \propto\ \exp(-\alpha\, k)$ which \textit{does not depend on the base amplitude}  $a$ for fixed values of other parameters;
  \item (\textit{d}) the increase of the spectral domain yields the increase of the cascade length but does not changes the exponent $\a$;
  \item (\textit{e}) the increase of the perturbation magnitude $\delta$ from $5\%$  to $50\%$ effects the time scale on which the MI occurs in the physical space: higher perturbation accelerates the process;
  \item (\textit{f}) the increase of the base wavenumber $k_0$ to $2k_0$, $3k_0,\ldots$, $30k_0$; changes the exponent $\a$ and yields at some point (beginning with $\approx 35k_0$) the occurrence of the inverse cascade;
\end{itemize}

\begin{figure}
  \centering
  \subfigure[$a = 0.08$, $t = 604$]{
  \includegraphics[width=7cm,height=6cm]{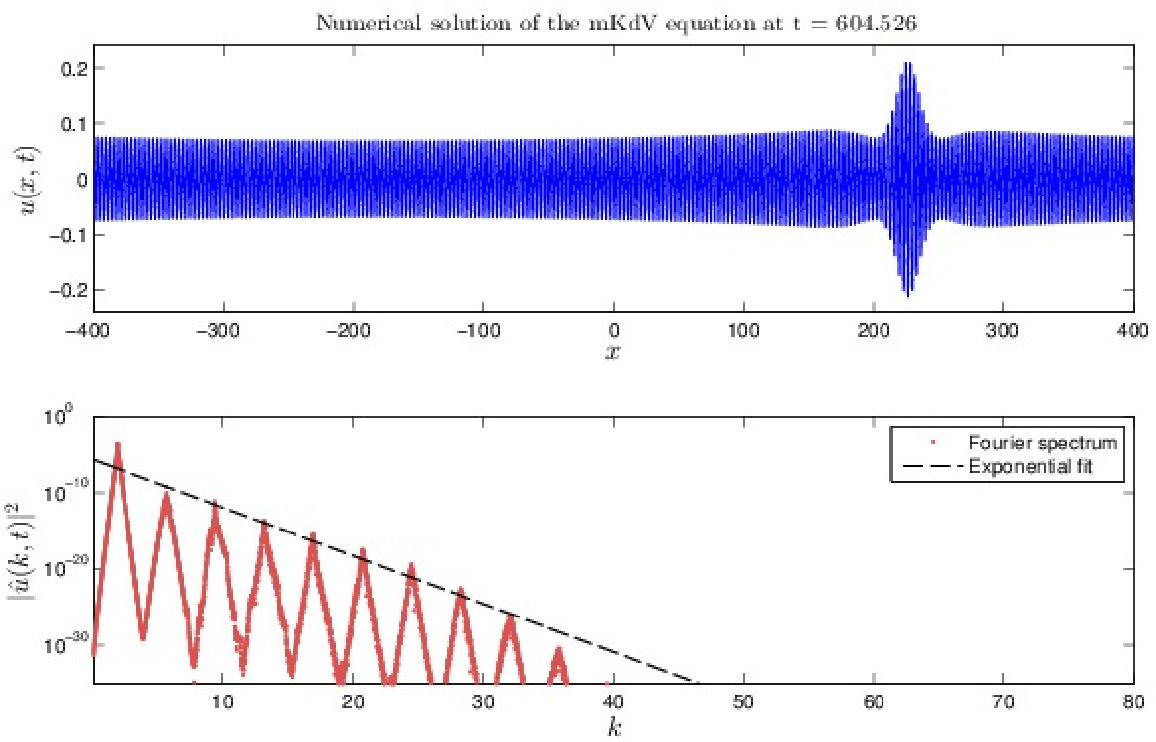}}
  \subfigure[$a = 0.16$, $t = 281$]{
  \includegraphics[width=7cm,height=6cm]{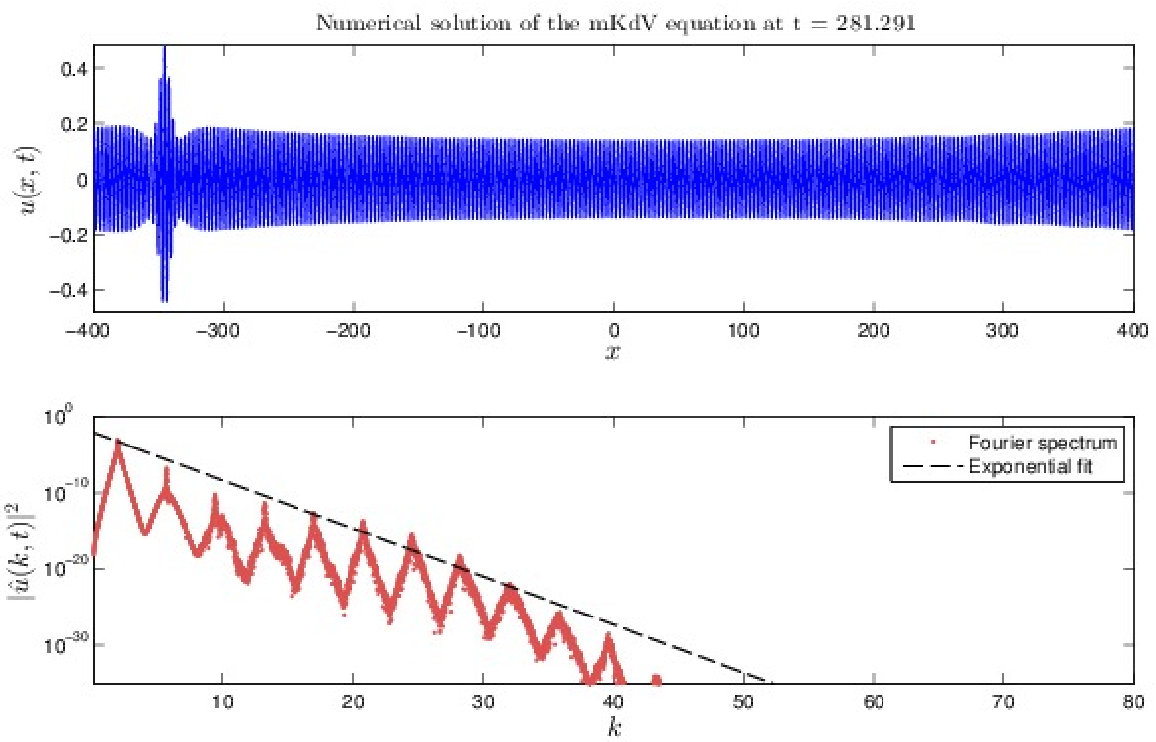}}
  \caption{\small\em Direct cascade in the mKdV equation. The energy spectrum exponential fit shown on bottom panels is the same for both amplitudes and equal to $\ue^{-1.45 k}$.}
  \label{fig:direct}
\end{figure}

The appearance of the direct $D$-cascade has been clearly observed (characteristic cascades are shown in Figure~\ref{fig:direct}), for the wide range of the initial parameters described in Table~\ref{tab:params}. For the sake of illustration, on Figure~\ref{fig:direct} we show the formation of the direct $D$-cascade for the initial condition \eqref{eq:pert} with the following parameters: $k_0 = 1.884$, $m = 0.05$, $K_0 = 0.00785$ and $a = 0.08$ on the top panel (\textit{a}) and $a = 0.16$ on the panel (\textit{b}). These values of parameters were also used in \cite{Grimshaw2005}. However, the numerical resolution was significantly lower in their study and the authors did not report the evolution of the system in the Fourier space. From our simulations it can be easily seen that an increase in the base wave amplitude does not change the shape of the energy spectrum along with the spacing between cascading modes. Moreover, the time instances $t = 50$ and $90$, where these snapshots were taken, indicate that the MI development goes faster with higher amplitudes. The dashed black lines shown on bottom panels show the same exponential fit for both panels.

In the next Section we present the results of our studies of the  inverse cascade, using the same parameters as above but with base wavenumbers $k_0$ shifted toward short waves.

\section{Double cascade in the mKdV equation}

In order to solve numerically the mKdV equation on a periodic domain we used the classical Fourier-type pseudo-spectral method along with the $2/3$-rule for dealising of nonlinear products. The discretization in time was done with the embedded adaptive 5$^{\,\mathrm{th}}$ order Cash--Karp Runge--Kutta scheme with the adaptive PI step size control. In most numerical simulations presented below we will use $N\ =\ 32\, 768$ Fourier harmonics, unless a special remark is made.

\begin{figure*}[hbtp!]
  \centering
  \subfigure[$t=5$]{
  \includegraphics[width=7cm,height=6cm]{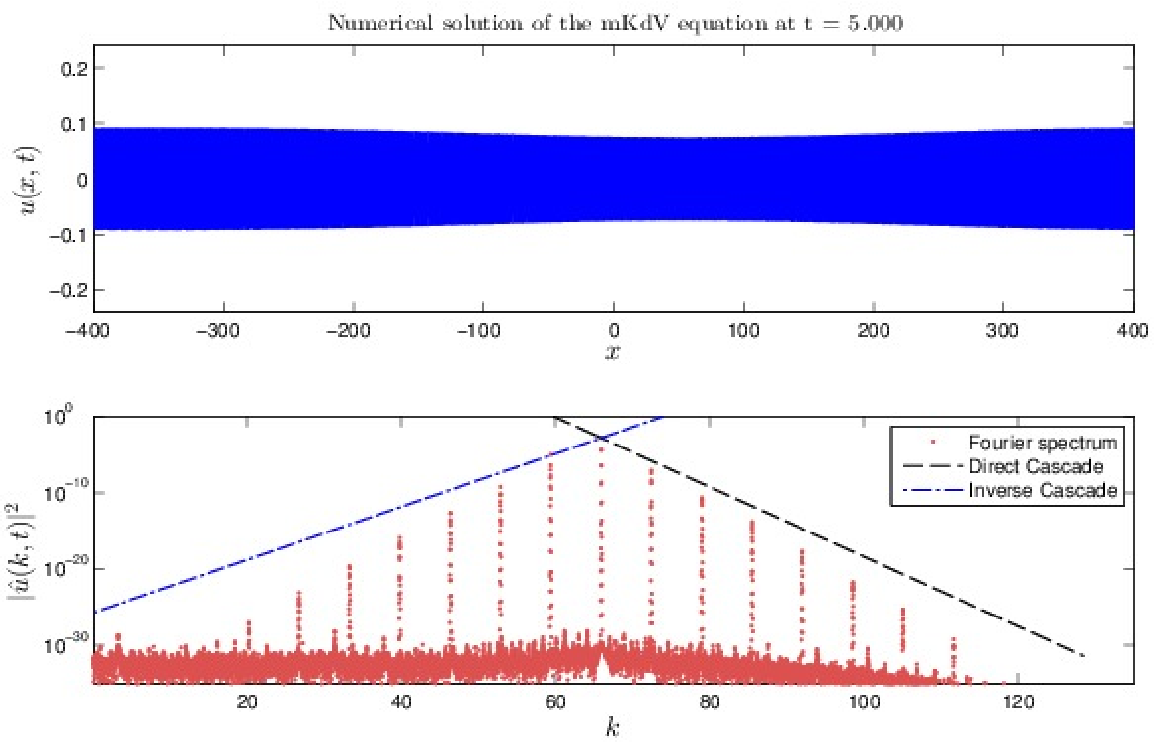}}
  \subfigure[$t=10$]{
  \includegraphics[width=7cm,height=6cm]{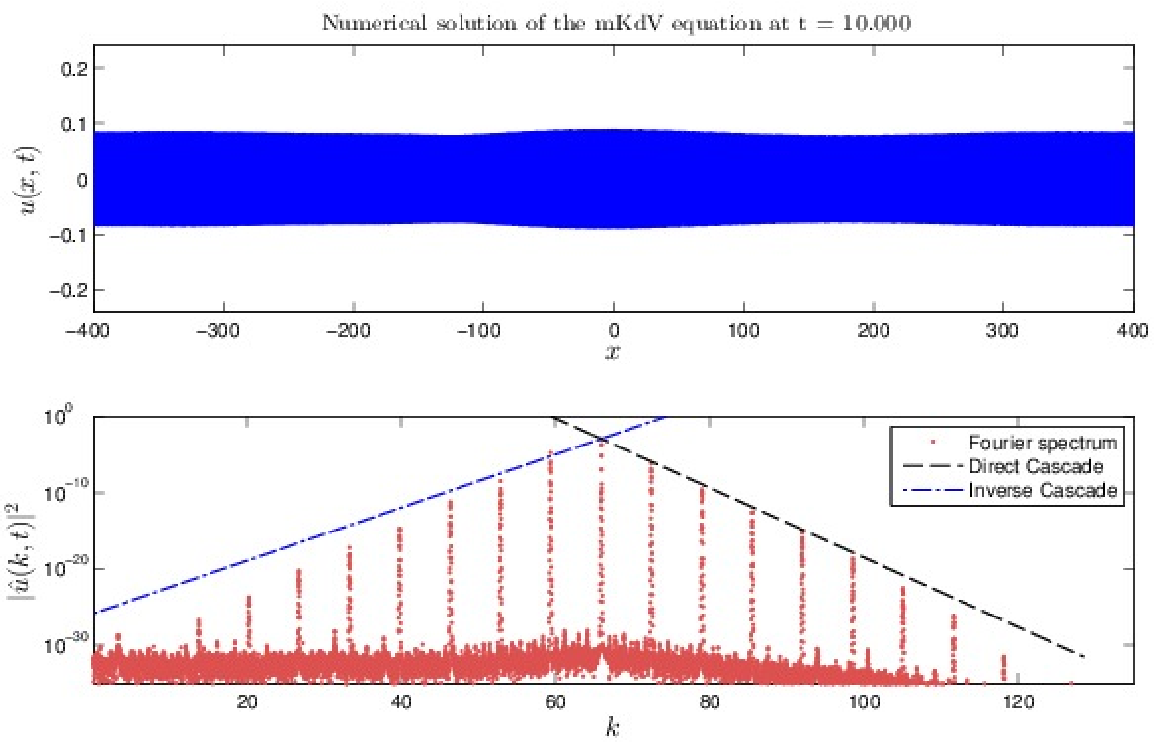}}
  \subfigure[$t=15$]{
  \includegraphics[width=7cm,height=6cm]{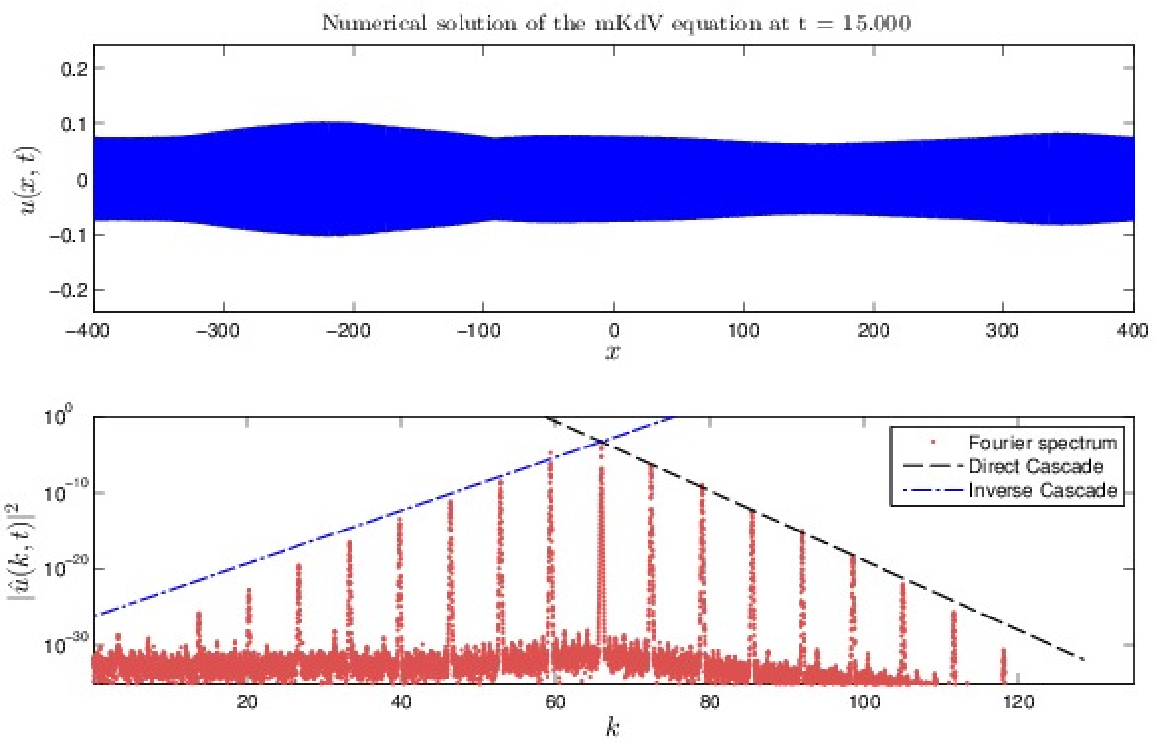}}
  \subfigure[$t=20$]{
  \includegraphics[width=7cm,height=6cm]{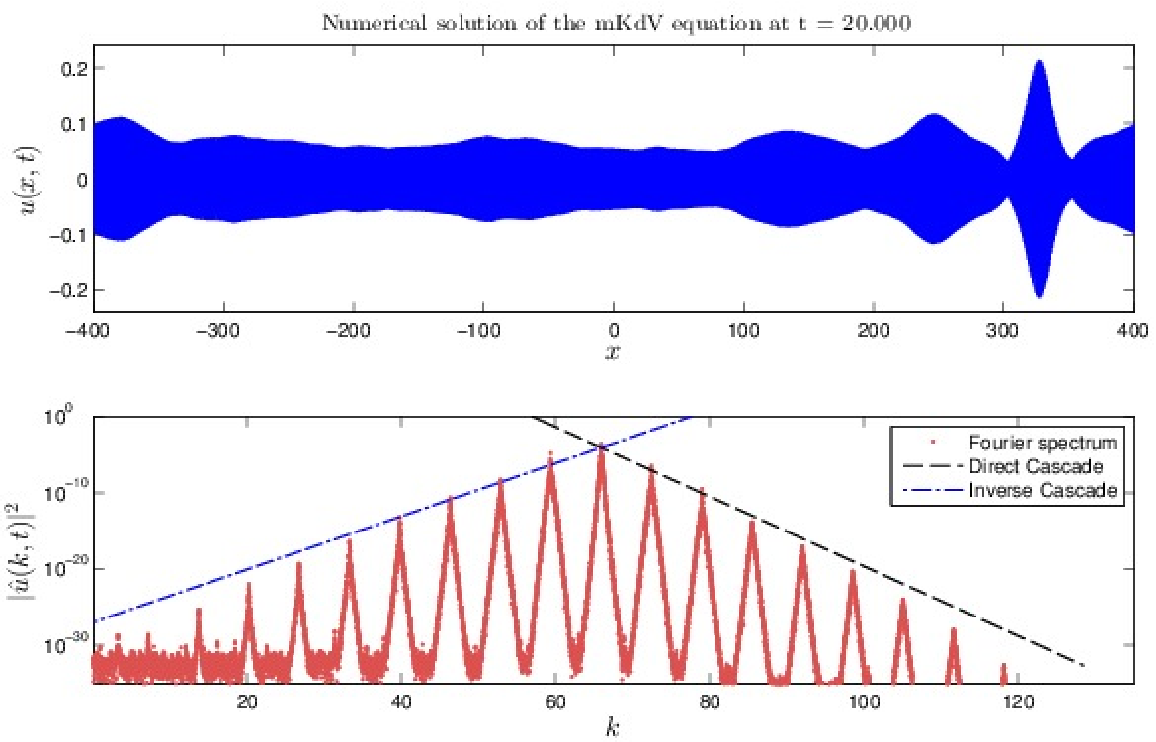}}
  \subfigure[$t=25$]{
  \includegraphics[width=7cm,height=6cm]{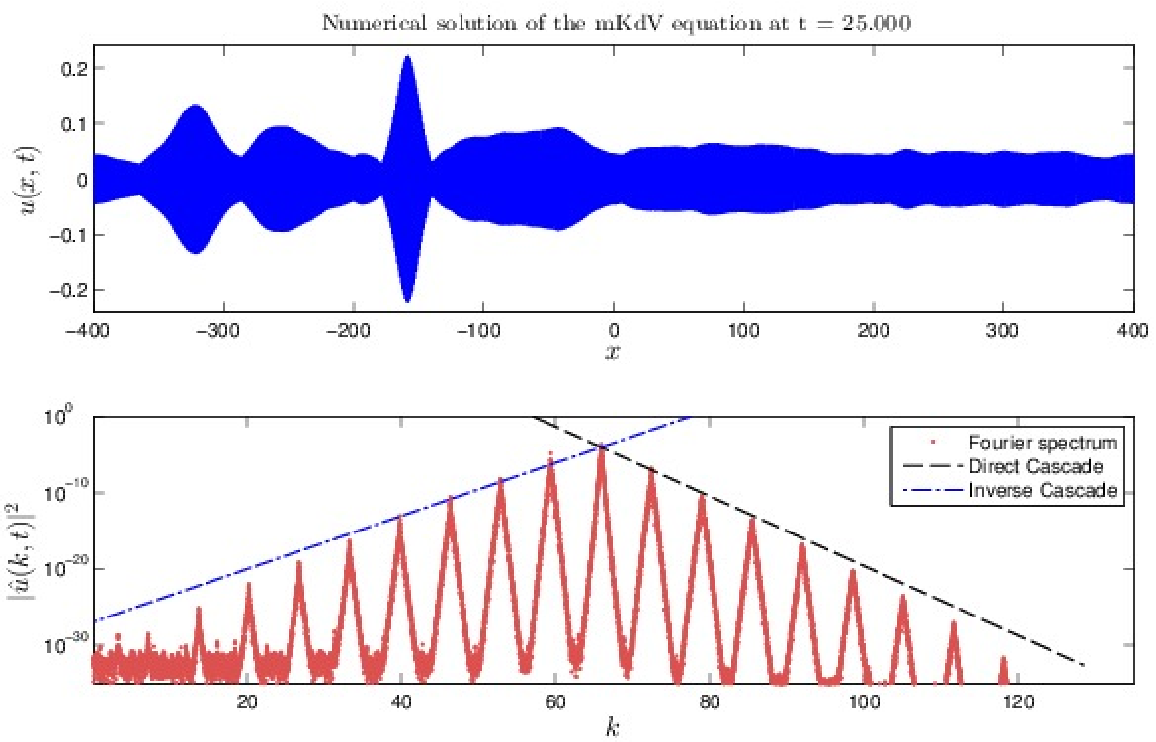}}
  \subfigure[$t=75$]{
  \includegraphics[width=7cm,height=6cm]{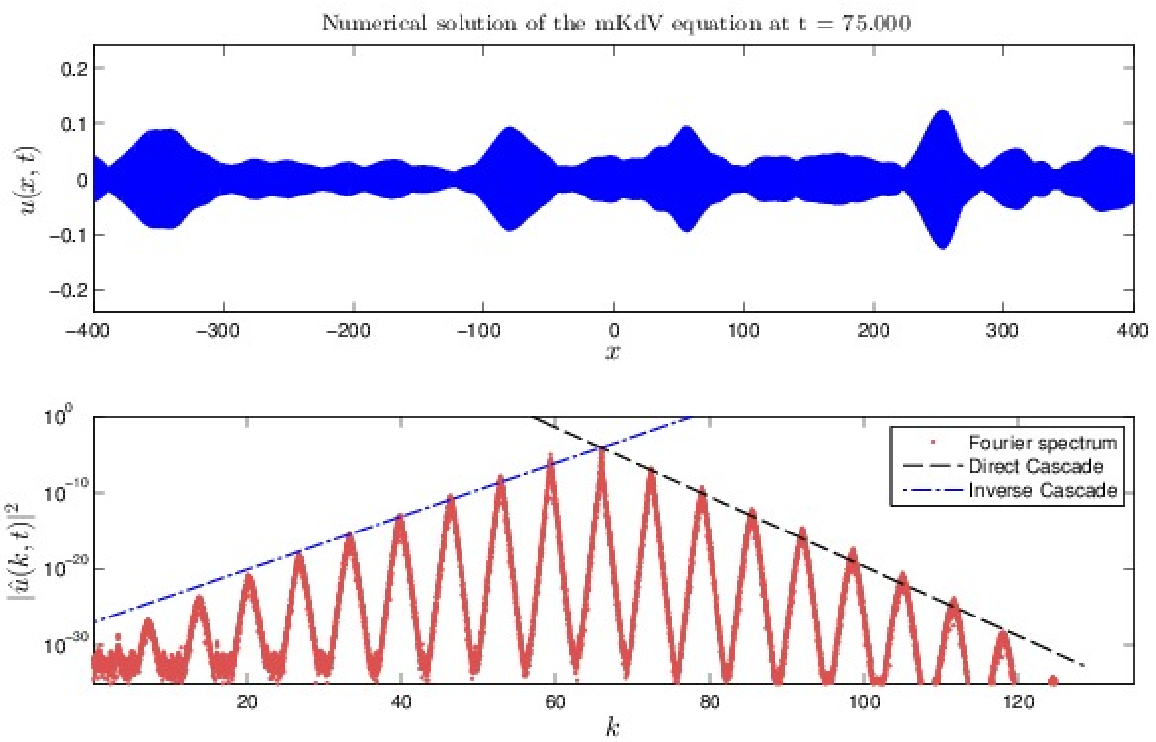}}
  \caption{\small\em Formation of the direct and inverse cascades in the mKdV equation.}
  \label{fig:ampl08}
\end{figure*}

On Figure~\ref{fig:ampl08} we show the simultaneous formation of the direct and inverse cascades (which will be referred below as the \emph{double cascade}) at different times. In order to achieve this result the base wavenumber $k_0$ was shifted approximatively into the middle of the spectral domain ($k_0 = 35\times 1.884$). One can see that starting from $t \approx 20$ (see Figure~\ref{fig:ampl08}(\textit{d})) the MI is already fully developed in the physical space, which corresponds to the appearance of a quasi-stationary structure -- double cascade in the Fourier space. For times $t\geq 20$ the evolution of the system follows the FPU-like pattern, without any significant change in the Fourier domain (see Figure~\ref{fig:ampl08}(\textit{e, f})).

As in the case of the isolated direct cascade, the double cascade equally has the exponential decay in the Fourier space. However, the two exponents are slightly different in a way that the direct cascade decays generally faster than its inverse counterpart.

\subsection{Effect of the amplitude}

\begin{figure}
  \centering
  \subfigure[$t = 5$]{\includegraphics[width=7cm,height=6cm]{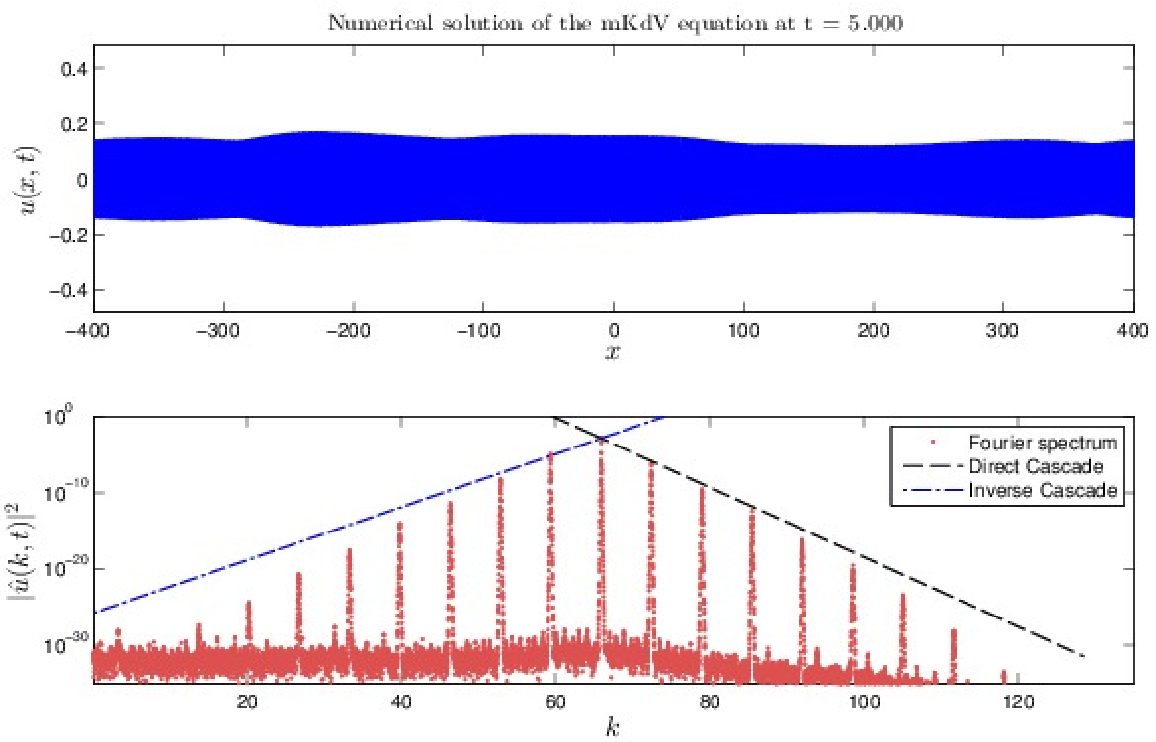}}
  \subfigure[$t = 10$]{\includegraphics[width=7cm,height=6cm]{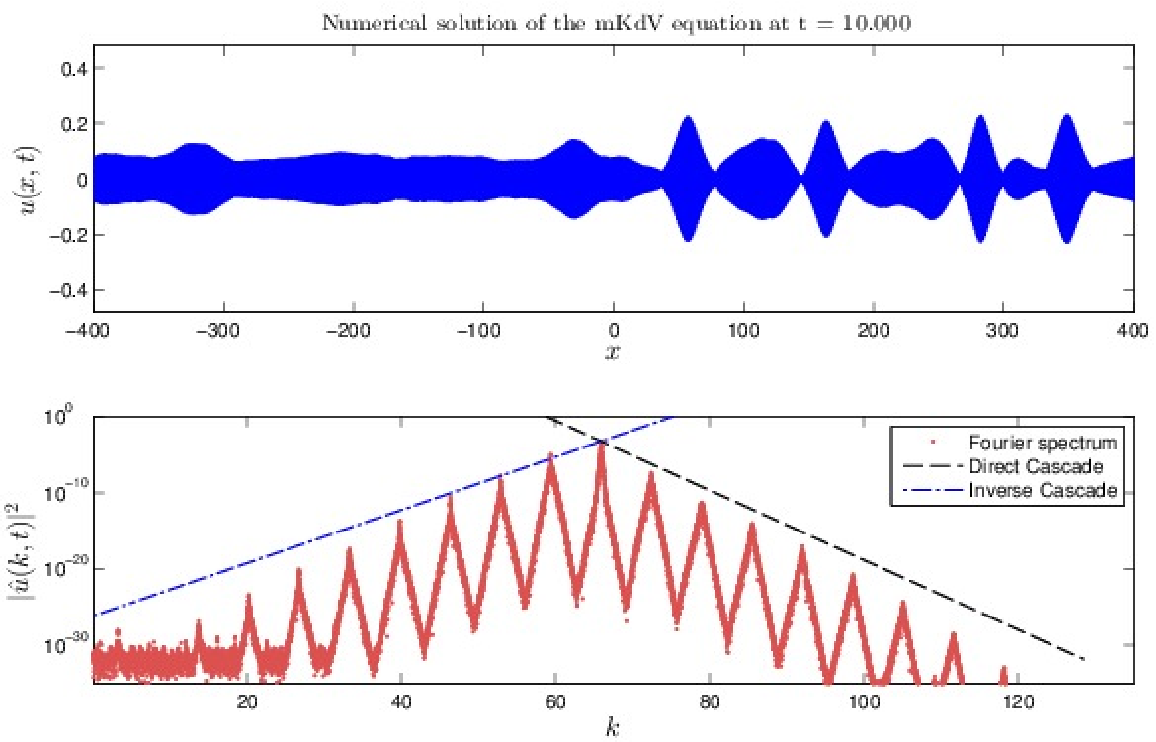}}
  \subfigure[$t = 30$]{\includegraphics[width=7cm,height=6cm]{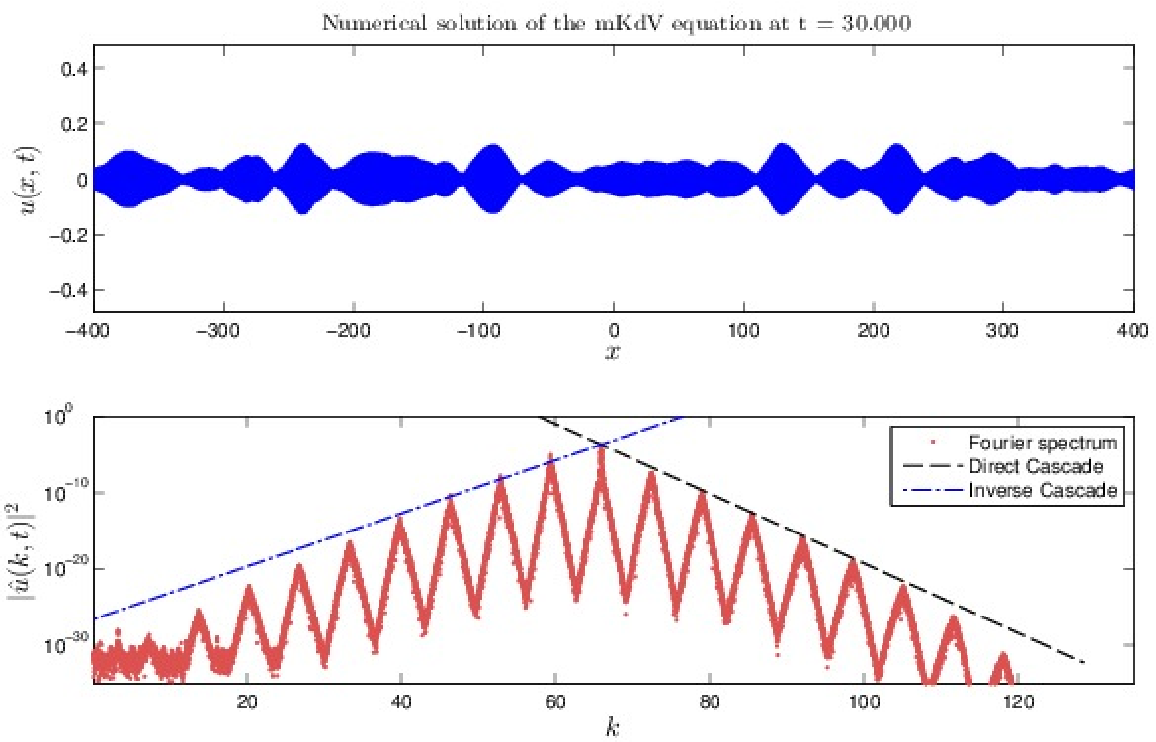}}
  \caption{\small\em Development of the double cascade for the initial wave amplitude $a = 2\times 0.08$.}
  \label{fig:ampl16}
\end{figure}

We performed the same simulation as shown on Figure~\ref{fig:ampl08}, but with twice the initial wave amplitude $a = 2\times 0.08$. The results are demonstrated on Figure~\ref{fig:ampl16}. One can see that the main characteristics of the double cascade remain unaffected, \textit{i.e.} the number and position of cascading modes along with the exponential shape of the Fourier spectrum. This observation can be made even more precise as we plot the same fitted exponents on bottom panels. One can see on Figure~\ref{fig:ampl16}(\textit{c}) a very good agreement between the observed spikes with the observed above exponential decay rate.

In contrast to the previous case with $a = 0.08$, all the processes are accelerated. Most importantly, the development of the MI in the physical space happens earlier with respect to the former computations shown on Figure~\ref{fig:ampl08}. Consequently, the process enters in the FPU-like recurrence regime earlier showing that the cascade formation is intrinsically linked to the MI development.


\subsection{Effect of the perturbation}

We studied also the effect of the perturbation magnitude $\delta$. We do not report the numerical results here due to the limitations in the Letter length. However, we report that for $m\ =\ 0.05\ \div\ 0.5$ (\textit{i.e.} 5\% to 50\% of the base wave amplitude) the same processes described above still take place. Basically, the higher values of the parameter $m$ yield the larger initial amount of the energy in the system which lead to the faster development of the MI.


\subsection{Effect of the spectral domain}

\begin{figure}
  \centering
  \subfigure[$t = 15$]{\includegraphics[width=7cm,height=6cm]{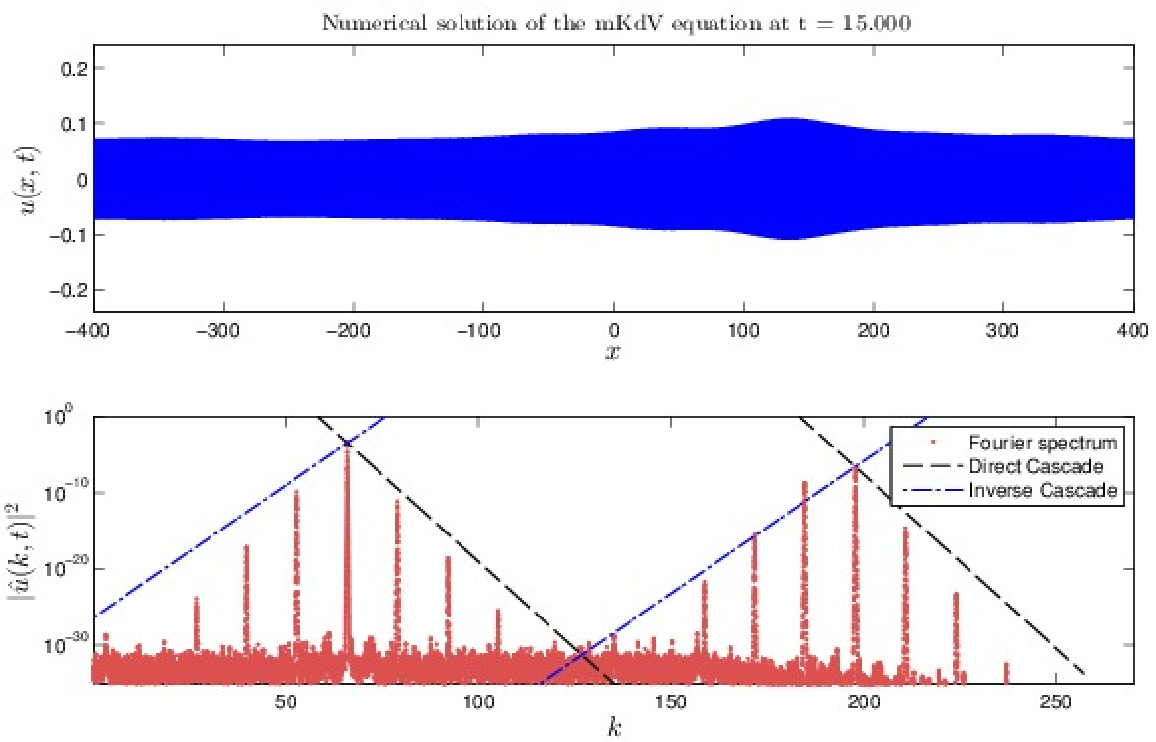}}
  \subfigure[$t = 20$]{\includegraphics[width=7cm,height=6cm]{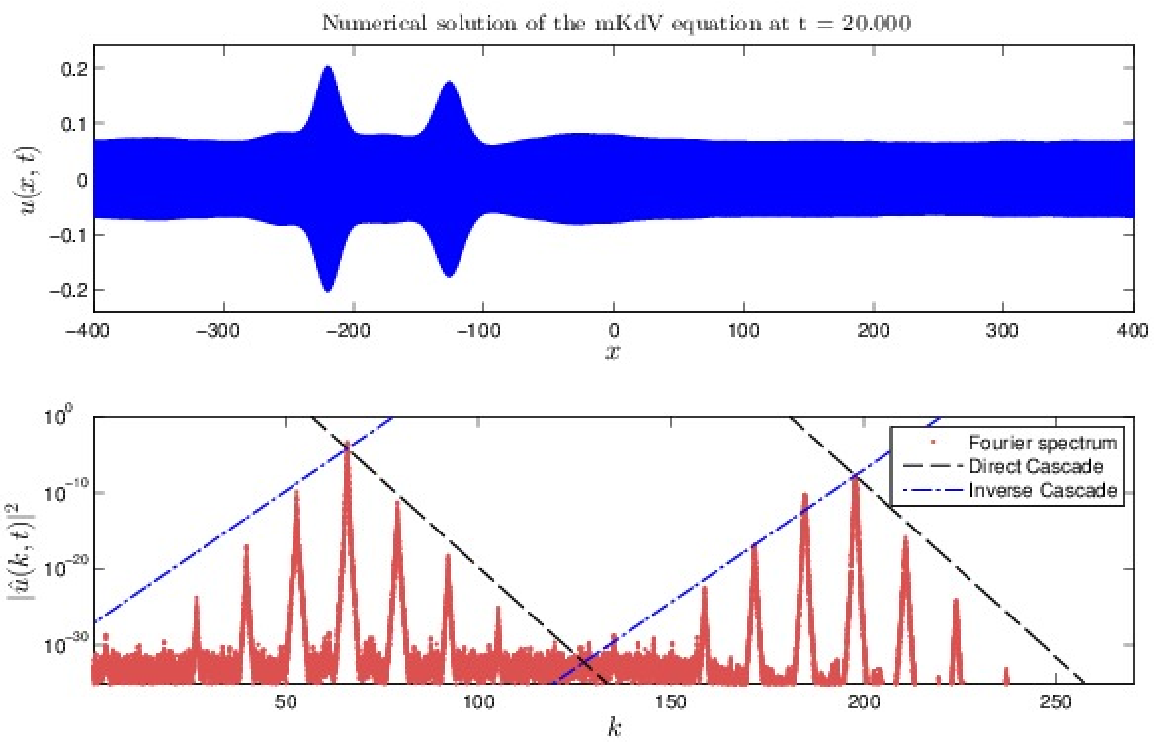}}
  \subfigure[$t = 25$]{\includegraphics[width=7cm,height=6cm]{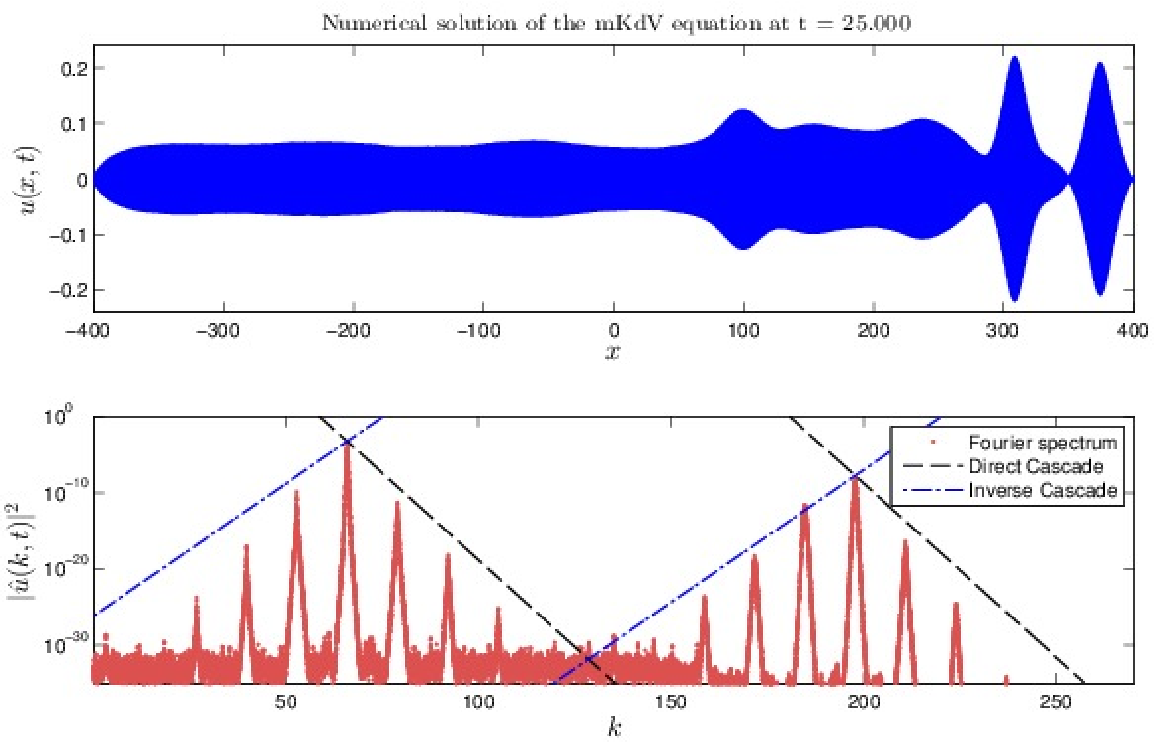}}
  \caption{\small\em Formation of two copies of the double inverse and direct cascades in a two times larger spectral domain. The exponents shown with dashed lines were measured in the simulations from Figure~\ref{fig:ampl08}. The observed discrepancy shows the effect of the spectral domain on the energy spectrum shape.}
  \label{fig:double}
\end{figure}

The last series of numerical experiments reported in this Letter aim at the studying the effect of the spectral domain on the energy cascade shape. With this purpose in mind we keep the same initial conditions as above, however we double the size of the spectral domain (consequently, the number of Fourier modes becomes $N\ =\ 65\, 536$ correspondingly). These results are shown on Figure~\ref{fig:double}. Our initial intuition, based on the experience with an isolated direct cascade, indicated that the increase of the spectral domain will lead to the change of the spectrum exponents by preserving the global structure of the double cascade.

Surprisingly, our simulations show that the cascade observed on Figure~\ref{fig:ampl08} preserves its shape and location in the Fourier space ($0\leq k < 120$), while in the additional space ($120 \leq k \leq 240$) another (apparently slightly smaller) replica of the initial double cascade appears, shifted farther in the Fourier domain.

Each replica is an equally stable structure in the Fourier space as the double cascade described above (\textit{i.e.} with evolution the number and location of cascading modes remain the same along with the spectrum exponents). However, they are different from the isolated double cascade observed for precisely the same initial conditions. The main difference consists, for example, in the number of cascading modes which obviously affects the exponent of the energy spectrum shape.

By comparing the development of the MI in the physical space on Figures~\ref{fig:ampl08}(\textit{d,e}) and \ref{fig:double}(\textit{b,c}) we can see that in the case of the smaller spectral domain the MI is more developed comparing to the latter. This phenomenon can be explained by the presence of two recurrent coherent structures (instead of one) each of them having a stabilizing effect on the MI.

\section{Summary}

Integrability of a specific evolutionary nonlinear PDE is a mathematical issue related to the spectrum of the corresponding Lax operator \cite{Ablowitz2001}. It gives a priori no information about the possible scenarios of the energy transport over the scales (in direct, inverse or both directions) as well as about the shape of the energy spectra. A good illustration is the nonlinear Schr\"odinger (NLS) equation which is integrable but this fact does not prevent numerous researchers from studying the corresponding energy spectra numerically, experimentally and analytically even nowadays \cite{Kuksin2013}. Let us recall that the complete integrability of the NLS equation was proved by Zakharov \& Shabat in 1972 \cite{Zakharov1972}, while the observation of the inverse cascade in the NLS was done only 20 years later \cite{Dyachenko1992}.

The energy transport in the mKdV, also known to be integrable by the inverse scattering transform, is even less studied than the NLS, and the existence of the inverse energy cascade in the mKdV is a novel phenomenon which we believe should be studied further on, both numerically and experimentally. In this Letter we presented some preliminary but very promising results on the formation of the inverse energy cascade in the mKdV equation. Briefly our findings can be summarized as:
\begin{itemize}
  \item Contrary to the direct $D$-cascade which can be observed for any $k_0 > 0$, the inverse cascade starts to appear only for large values of $k_0$.
  \item It is not possible to observe an isolated inverse $D$-cascade, since the direct cascade forms for any choice of $k_0$. As a result, our observations show the so-called double cascade which is composed of at least one instance of the direct and inverse cascades.
  \item The exponents of the inverse and direct cascades are slightly different. In general, the direct cascade tends to decay faster. Moreover, the values of the exponents were shown not to depend on the initial wave amplitude, but only on the initial wavenumber $k_0$.
  \item The double cascade appears to be a very robust quasi-stationary structure in the Fourier space. In other words, the system at this stage enters into the FPU-like recurrent regime.
  \item The formation of the $D$-cascade is accompanied by the nonlinear development of the MI.
  \item The increase of the spectral domain yields the formation of two (and possibly even more) replica of the double cascade, each containing roughly one half of the initial energy. Our conjecture is that the further increase in the spectral domain will lead to the appearance of more and more replica of this double cascade structure. Or in other words, we will see the appearance of more and more recurrent structures sharing the total system energy. This process may be regarded as a mechanism of the MI stabilization.
\end{itemize}

The problem of the MI stabilization was considered by several research groups \cite{Segur2005} and the main stabilization mechanism which was put forward has been related to some sort of dissipation. In this study we propose another mechanism which can work in fully conservative systems.

The present investigation is restricted to the case of moderate nonlinearities ($a\ \sim\ \O\bigl(10^{-1}\bigr)$), since in this case it is straightforward to find an unstable mode $k_0$. When we go to the field of the finite amplitude waves ($a\ \sim\ \O(1)$), the stability study becomes a complicated mathematical problem \cite{Johnson2010}. The investigation of the $D$-cascade formation in strongly nonlinear regimes, occuring for example in plasma physics \cite{Ruderman2008}, is among our next priorities.

\acknowledgments

This research has been supported by the Austrian Science Foundation (FWF) under projects P22943-N18 and P24671 as well as by the National Science Foundation under Grant No. NSF PHY11-25915.

\end{document}